\documentclass[showpacs]{revtex4}
\usepackage{epsfig}
\def\beq{\begin{equation}}
\def\enq{\end{equation}}
\def\beqa{\begin{eqnarray}}
\def\enqa{\end{eqnarray}}

\def\MeV{\nobreak\,\mbox{MeV}}
\def\GeV{\nobreak\,\mbox{GeV}}

\def\qq{\lag\bar{q}q\rag}

\def\ss{\lag\bar{s}s\rag}
\def\mix{\lag\bar{q}g\si.Gq\rag}

\def\Gd{\lag g^2G^2\rag}
\def\G3{\lag g^3G^3\rag}

\def\rh{\rho}
\def\si{\sigma}

\def\al{\alpha}
\def\be{\beta}
\def\alma{\alpha_{max}}
\def\almi{\alpha_{min}}
\def\bemi{\beta_{min}}
\def\lb{\label}
\def\nn{\nonumber}

\newcommand{\rag}{\rangle}
\newcommand{\lag}{\langle}

\begin{document}

\title{\sc
QCD sum rules study of the meson $Z^+(4430)$
}
\author{Su Houng Lee}
\email{suhoung@phya.yonsei.ac.kr}
\affiliation{Institute of Physics and Applied Physics, Yonsei University,
Seoul 120-749, Korea}
\author{Antonio Mihara}
\email{mihara74@gmail.com}
\affiliation{Instituto de Ciências Exatas e Tecnologia, Universidade Federal
do Amazonas,
R. Nossa Senhora do Rosário 3863, 69100-000 - Itacoatiara, AM, Brazil}
\author{Fernando S. Navarra}
\email{navarra@if.usp.br}
\affiliation{Instituto de F\'{\i}sica, Universidade de S\~{a}o Paulo,
C.P. 66318, 05389-970 S\~{a}o Paulo, SP, Brazil}
\author{Marina Nielsen}
\email{mnielsen@if.usp.br}
\affiliation{Instituto de F\'{\i}sica, Universidade de S\~{a}o Paulo,
C.P. 66318, 05389-970 S\~{a}o Paulo, SP, Brazil}

\begin{abstract}
We use QCD  sum rules to study the recently observed meson $Z^+(4430)$,
considered as a $D^*D_1$ molecule with $J^{P}=0^{-}$.
We consider the contributions of condensates up to dimension eight and
work at leading order in $\alpha_s$. We get $m_Z=(4.40\pm0.10)~\GeV$ in a very
good agreement with the experimental value. We also make predictions for the
analogous mesons $Z_{s}$ and $Z_{bb}$ considered as $D_s^*D_1$ and
$B^*B_1$ molecules respectively. For $Z_{s}$
we predict $m_{Z_{s}}= (4.70\pm 0.06)~{\rm GeV}$, which is above the
$D_s^*D_1$ threshold, indicating that it is probably a very broad state and,
therefore, difficult to observe experimentally. For $Z_{bb}$ we predict
$m_{Z_{bb}}= (10.74\pm 0.12)~{\rm GeV}$, in agreement with quark
model predictions.
\end{abstract}

\maketitle


In the last years many new mesons have been observed by the BaBar, BELLE, CLEO,
D0 and FOCUS collaborations. Among these new mesons, some have been considered
as good candidates for tetraquark states like the $D_{sJ}(2317)$ \cite{babar1},
the $D_{sJ}(2460)$ \cite{cleo1}, the $X(3872)$ \cite{belle1} and more
recently the $Z^+(4430)$ \cite{belle2}. While there are many indications
that the charmed mesons, $D_{sJ}(2317)$ and $D_{sJ}(2460)$, are not four-quark
states \cite{swan,tetra}, this is not the case for the charmonium like states,
$X(3872)$ and $Z^+(4430)$. The $X(3872)$, with quantum numbers
$J^{PC} = 1^{++}$, does not fit in the charmonium spectrum and presents a
strong isospin violating decay, disfavoring a $c \bar{c}$ assignment.
The $Z^+(4430)$, recently observed in the $Z^+\to \psi^\prime\pi^+$ decay mode
\cite{belle2}, is the most interesting one since, being a charged state, it
can not be a pure $c\bar{c}$ state.

There are already many theoretical interpretations for the structure of the
$Z^+(4430)$ meson: molecular $D^*D_1$ state \cite{meng}, tetraquark
state \cite{maiani,rosner}, or a cusp in the $D^*D_1$ channel \cite{bugg}.
In ref. \cite{maiani},
the authors have interpreted the  $Z^+(4430)$ meson
as the first radial excitation of the diquark-antidiquark
$[cu][\bar{c}\bar{d}]$ state, with $J^{PC}=1^{+-}$. The
low lying tetraquark state, $[cu][\bar{c}\bar{d}]$, is interpreted as the
charged partner of the $X(3872)$ meson. Supposing that the mass of the
low lying tetraquark state, $[cu][\bar{c}\bar{d}]$ is compatible with the mass
of the $X(3872)$ meson, the mass difference between the $Z^+(4430)$ and
the $X(3872)$  would be close to the mass difference between the $\psi^\prime$
and $J/\psi$: $m_{\psi'}-m_{J/\psi}=590~\MeV$. Therefore, they arrive at
$m_{Z}\sim 3872+590\sim 4460\MeV$, which is compatible with the observed mass.
In ref. \cite{meng}, the closeness of the $Z^+(4430)$ mass to the threshold
of $D^{*+}(2010)\bar{D}_1(2420)$ lead the authors to consider the $Z^+(4430)$
as a $D^*\bar{D}_1$ molecule. In this case, the allowed $J^P$ of $Z$ would
be $0^-,~1^-$ or $2^-$, although the $2^-$ assignment is probably suppressed
in the $B\to Z(4430)K$ decay, by the small phase space.   Among the remaining
possible $0^-$ and $1^-$ states, the former will be more stable as the later 
can  also decay to $D\bar{D}_1$ in s-wave.  Moreover, one expects a bigger 
mass for a $J^P=1^-$ state as compared to a $J^P=0^-$ state.   Therefore,
in this work we use QCD sum rules (QCDSR) \cite{svz,rry,SNB}, to
study the two-point function of the state $Z^+(4430)$ considered as a
$D^*D_1$ molecule with $J^P=0^-$.

In a previous calculation, the QCDSR approach was used to study
the $X(3872)$ meson, considered as a diquark-antidiquark
state, and a good agreement with the experimental mass was obtained 
\cite{x3872}.
If we suppose, as in ref.~\cite{maiani}, that the $Z^+(4430)$
is related to the first radial excitation of the $X(3872)$, in the QCDSR
approach its mass would be given by $\sqrt{s_0}$, where $s_0$ is the continuum
threshold. In ref.~\cite{tetra} it was found that $\sqrt{s_0}=(4.3\pm0.1)~\GeV$
also
in a very good agreement with the experimental mass of $Z^+(4430)$. However,
this is not a precise determination of the mass of the first excited state,
since the continuum threshold gives only a lower bound for the mass of the
first excited states.


Considering $Z^+(4430)$ as a $D^*D_1$ molecule with $J^P=0^-$, a possible
current describing such state is given by:
\beq
j={1\over\sqrt{2}}\left[(\bar{d}_a\gamma_\mu c_a)(\bar{c}_b\gamma^\mu\gamma_5
u_b)+(\bar{d}_a\gamma_\mu\gamma_5 c_a)(\bar{c}_b\gamma^\mu u_b)\right]\;,
\label{field}
\enq
where $a$ and $b$ are color indices, We have considered the symmetrical
state $D^{*+}\bar{D}_1^0+\bar{D}^{*0}D_1^+$ because it has positive $G$-parity,
which is consistent with the observed decay $Z^+(4430)\to\psi^\prime\pi^+$.

The two-point correlation function is given by:
\beq
\Pi(q)=i\int d^4x ~e^{iq.x}\lag 0
|T[j(x)j^\dagger(0)]|0\rag.
\lb{2po}
\enq

On the OPE side, we work at leading order in $\alpha_s$ and consider the
contributions of condensates up to dimension eight.  We calculate
the light quark part of the correlation function
in the coordinate-space and we use the momentum-space expression for the
charm quark propagator. The resulting light-quark part is then Fourier
transformed to the momentum space in $D$ dimensions and it is dimensionally
regularized at $D=4$.

The correlation function in the OPE side can be written as a
dispersion relation:
\beq
\Pi^{OPE}(q^2)=\int_{4m_c^2}^\infty ds {\rho^{OPE}(s)\over s-q^2}
+\Pi^{mix\qq}(q^2)\;,
\lb{ope}
\enq
where $\rho^{OPE}(s)$ is given by the imaginary part of the
correlation function: $\pi \rho^{OPE}(s)=\mbox{Im}[\Pi^{OPE}(s)]$.
We get:
\beq
\rho^{OPE}(s)=\rho^{pert}(s)+\rh^{\qq}(s)+\rh^{\lag G^2\rag}(s)
+\rh^{mix}(s)+\rh^{\qq^2}(s)\;,
\lb{rhoeq}
\enq
with
\beqa\label{eq:pert}
&&\rho^{pert}(s)={3\over 2^{9} \pi^6}\int\limits_{\almi}^{\alma}
{d\al\over\alpha^3}
\int\limits_{\bemi}^{1-\al}{d\be\over\be^3}(1-\al-\be)
\left[(\al+\be)m_c^2-\al\be s\right]^4,
\nn\\
&&\rho^{\qq}(s)=0,
\nn\\
&&\rho^{\lag G^2\rag}(s)={m_c^2\Gd\over2^{8}\pi^6}
\int\limits_{\almi}^{\alma} {d\al\over\al^3}
\int\limits_{\bemi}^{1-\al}d\be(1-\al-\be)\left[(\al+\be)m_c^2-\al\be s\right],
\nn\\
&&\rho^{mix}(s)=0,
\nn\\
&&\rho^{\qq^2}(s)=-{m_c^2\qq^2\over 4\pi^2}\sqrt{1-4m_c^2/s},
\nn\\
&&\Pi^{mix\qq}(q^2)={m_c^2\mix\qq\over 2^3\pi^2}\int_0^1
d\al{\al(1-\al)\over m_c^2-\al(1-\al)q^2}\left[1+{m_c^2
\over m_c^2-\al(1-\al)q^2}\right].
\label{dim8}
\enqa
where the integration limits are given by $\almi=({1-\sqrt{1-
4m_c^2/s})/2}$, $\alma=({1+\sqrt{1-4m_c^2/s})/2}$ and $\bemi={\al
m_c^2/( s\al-m_c^2)}$.
The contribution of dimension-six condensates $\lag g^3 G^3\rag$
is neglected, since it is suppressed by   the loop factor $1/16\pi^2$.
In Eq.~(\ref{ope}) the $\Pi^{mix\qq}(q^2)$ term is treated separately because
its imaginary part is proportional to delta functions and, therefore, can be 
easily integrated \cite{hen}.

It is very interesting to notice that the current in Eq.~(\ref{field}) does
not get contribution from the quark and mixed condensates. This is very 
different
from the OPE behavior obtained to the scalar-diquark axial-antidiquark current
used for the $X(3872)$ meson in ref.~\cite{x3872}, but very similar to the OPE
behavior obtained for the axial double-charmed meson $T_{cc}$, also
described by a scalar-diquark axial-antidiquark current \cite{tcc}.

The calculation of the phenomenological side at the
hadron level proceeds by writing a dispersion relation to the
correlation function in Eq.~(\ref{2po}):
\beq
\Pi^{phen}(q^2)=\int ds\, {\rho^{phen}(s)\over s-q^2}\,+\,\cdots\,,
\label{phen}
\enq
where $\rho^{phen}(s)$ is the spectral density and the dots represent
subtraction terms. The spectral density is described, as usual, as a single 
sharp
pole representing the lowest resonance plus a smooth continuum representing
higher mass states:
\beqa
\rho^{phen}(s)&=&f_Z^2\delta(s-m_{Z}^2) +\rho^{cont}(s)\,,
\label{den}
\enqa
where $f_Z$ gives the coupling of the current to the meson $Z^+$:
\beq\label{eq: decay}
\lag 0 |
j|Z^+\rag =f_Z.
\enq

For simplicity, it is
assumed that the continuum contribution to the spectral density,
$\rho^{cont}(s)$ in Eq.~(\ref{den}), vanishes bellow a certain continuum
threshold $s_0$. Above this threshold, it is assumed to be given by
the result obtained with the OPE. Therefore, one uses the ansatz \cite{io1}
\beq
\rho^{cont}(s)=\rho^{OPE}(s)\Theta(s-s_0)\;,
\enq

 After making a Borel transform to both sides of the sum rule, and
transferring the continuum contribution to the OPE side, the sum rules
for the pseudoscalar meson $Z^+$, up to dimension-eight condensates, can
be written as:
\beq f_Z^2e^{-m_Z^2/M^2}=\int_{4m_c^2}^{s_0}ds~
e^{-s/M^2}~\rho^{OPE}(s)\; +\Pi^{mix\qq}(M^2)\;, \lb{sr}
\enq
where
\beqa
\Pi^{mix\qq}(M^2)={m_c^2\mix\qq\over 8\pi^2}\int_0^1
d\al\,\exp\!\left[{-{m_c^2
\over\al(1-\al)M^2}}\right]\bigg[1+{m_c^2
\over\al(1-\al) M^2}\bigg]\,.
\label{8m2}
\enqa

To extract the mass $m_Z$ we take the derivative of Eq.~(\ref{sr})
with respect to $1/M^2$, and divide the result by Eq.~(\ref{sr}).

The values used for the quark
masses and condensates are \cite{SNB,narpdg}:
$m_c(m_c)=(1.23\pm 0.05)\,\GeV $,
$\lag\bar{q}q\rag=\,-(0.23\pm0.03)^3\,\GeV^3$,
$\lag\bar{q}g\si.Gq\rag=m_0^2\lag\bar{q}q\rag$ with $m_0^2=0.8\,\GeV^2$,
$\lag g^2G^2\rag=0.88~\GeV^4$.

We evaluate the sum rules in the Borel range $2.2 \leq M^2 \leq 3.5\GeV^2$,
and in the $s_0$ range $4.8\leq \sqrt{s_0} \leq5.0$ GeV.

\begin{figure}[h]
\centerline{\epsfig{figure=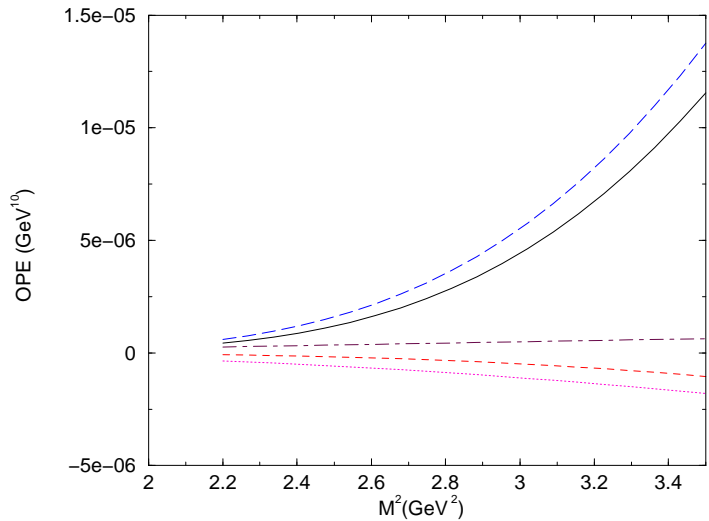,height=70mm}}
\caption{The OPE convergence in the region $2.2 \leq M^2 \leq
3.5~\GeV^2$ for $\sqrt{s_0} = 4.9$ GeV. Perturbative
contribution (long-dashed line), $\langle g^2G^2\rangle$ contribution
(dashed line), $\langle \bar{q}q\rangle^2$ contribution (dotted-line),
$\mix\qq$ (dot-dashed line) and the total contribution (solid line).}
\label{figconvtcc}
\end{figure}

From Fig.~\ref{figconvtcc} we see that we obtain a quite good OPE
convergence for $M^2\geq 2.5$ GeV$^2$. Therefore, we  fix the lower
value of $M^2$ in the sum rule window as $M^2_{min} = 2.5$ GeV$^2$.
This figure also
shows that, although there is a change of sign between
dimension-six and dimension-eight condensate contributions, as noticed in
\cite{oganes}, both contributions  are very small and, therefore, they do
not spoil the convergence of the OPE. It is also important to mention that
the OPE convergence in this case is much better than the OPE convergence for 
the
$X(3872)$ meson \cite{x3872}, and is comparable with the OPE convergence for
heavy baryons \cite{dunga}.

\begin{figure}[h]
\centerline{\epsfig{figure=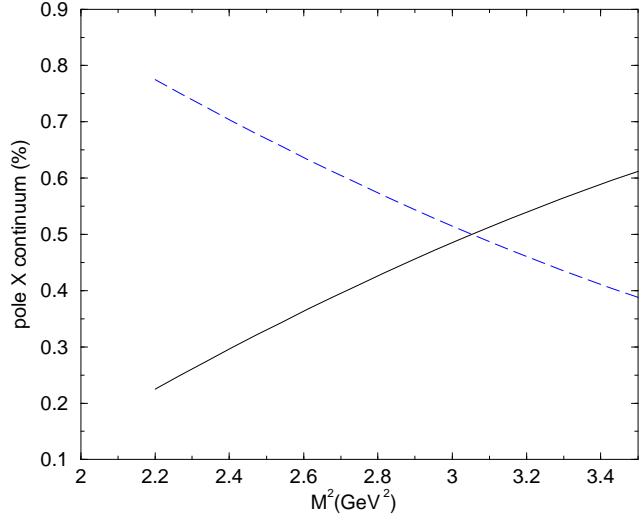,height=70mm}}
\caption{The dashed line shows the relative pole contribution (the
pole contribution divided by the total, pole plus continuum,
contribution) and the solid line shows the relative continuum
contribution for $\sqrt{s_0}=4.9~\GeV$.}
\label{figpvc}
\end{figure}

To get an upper limit constraint for $M^2$ we impose that
the QCD continuum contribution should be smaller than the pole contribution.
The comparison between pole and
continuum contributions for $\sqrt{s_0} = 4.9$ GeV is shown in
Fig.~\ref{figpvc}. From this figure we see that the pole contribution
is bigger than the continuum for $M^2\leq3.05~\GeV^2$.
The maximum value of $M^2$ for which this constraint is satisfied
depends on the value of $s_0$. The same analysis for the other values of the
continuum threshold gives $M^2 \leq 2.85$  GeV$^2$ for $\sqrt{s_0} = 4.8~\GeV$
and $M^2 \leq 3.25$  GeV$^2$ for $\sqrt{s_0} = 5.0~\GeV$.
In our numerical analysis, we shall then consider the range of $M^2$ values
from 2.5 $\GeV^2$ until the one allowed by the pole dominance criteria given
above.

\begin{figure}[h]
\centerline{\epsfig{figure=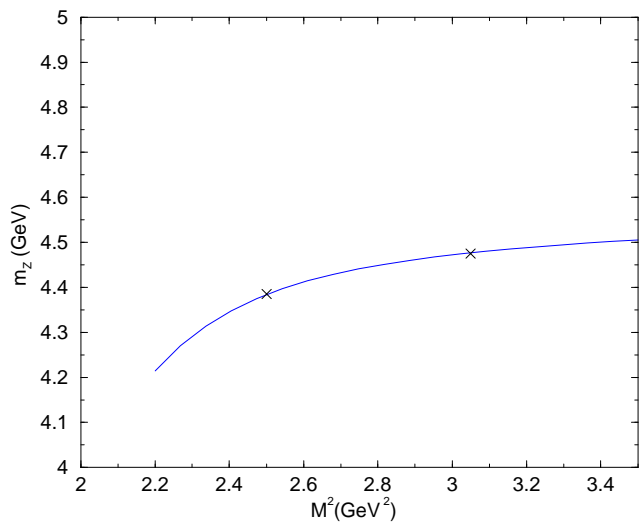,height=70mm}}
\caption{The $Z^+$ meson mass as a function of the sum rule parameter
($M^2$) for $\sqrt{s_0} =4.9$ GeV. The crosses
indicate the region allowed for the sum rules: the lower limit
(cut below 2.5 GeV$^2$) is given by OPE convergence requirement and the
upper limit by the dominance of the QCD pole contribution.}
\label{figmx}
\end{figure}
In Fig.~\ref{figmx}, we show the $Z^+$ meson mass, for $\sqrt{s_0} =
4.9~\GeV$, in the relevant sum rule window, with the upper and lower validity
limits indicated.  From this figure we see that
the results are reasonably stable as a function of $M^2$.

Using the Borel window, for each value of $s_0$, to evaluate the mass of the
$Z^+$ meson and then varying the value of the continuum threshold in the range
$\sqrt{s_0}=(4.9\pm0.1)~\GeV$, we arrive at
\beq
m_{Z} = (4.40\pm0.10)~\GeV,
\label{zmass}
\enq
in a very good agreement with the experimental value
\cite{belle2}.

To check the dependence of our results with the value of the
charm quark mass, we fix $\sqrt{s_0}=4.9~\GeV$ and vary the charm quark mass
in the range $m_c=(1.23\pm0.05)~\GeV$. Using $2.5\leq M^2\leq 3.05~\GeV^2$
we get: $m_{Z} = (4.43\pm0.08)~\GeV$, in agreement
with the result in Eq.~(\ref{zmass}). Therefore, we conclude that the most
important sources of uncertainty in our calculation is the value
of the continuum threshod and the Borel interval.

We can extend our results to the bottonium analogous state $Z_{bb}$,
considered as a pseudoscalar $B^*B_1$ molecule, by
exchanging the charm quark  in Eqs.~(\ref{field}) to (\ref{8m2}), by the bottom
quark. Therefore, in the case of the pseudoscalar meson $Z_{bb}$,
using consistently the perturbative $\overline{MS}$-mass $m_b(m_b)=(4.24
\pm0.6)~\GeV$, and the continuum threshold in the range
$11.2\leq\sqrt{s_0}\leq11.6~\GeV$,
we find a good OPE convergence for $M^2\geq8.0~\GeV^2$. The OPE convergence
in this case is even better than the one presented in Fig.~1.
We also find that the pole contribution is bigger than the continuum
contribution for $M^2\leq8.25~\GeV^2$ for $\sqrt{s_0}<11.2~\GeV$, and
for $M^2\leq9.9~\GeV^2$ for $\sqrt{s_0}<11.6~\GeV$. For $\sqrt{s_0}<11.2~\GeV$
we found no Borel window, since $M^2_{max}<8.0~\GeV^2$.

We find that the results for  the $Z_{bb}$ meson mass, in the allowed sum rule
window, are very stable as a function of $M^2$.
Taking into account the
variation of $M^2$ and varying $s_0$ and $m_b$ in the regions indicated we get:
\beq
\lb{masszb}
m_{Z_{bb}}=  (10.74\pm0.12)~\GeV~,
\enq
in a very good agreement with the prediction in ref.~\cite{cheung}.

For completeness, we also predict the mass of the strange analogous
meson $Z_{s}^+$ considered as a pseudoscalar $D_s^*D_1$ molecule. The current 
is obtained by exchanging the $d$ quark in Eq.~(\ref{field}) by the $s$ quark.

\begin{figure}[h]
\centerline{\epsfig{figure=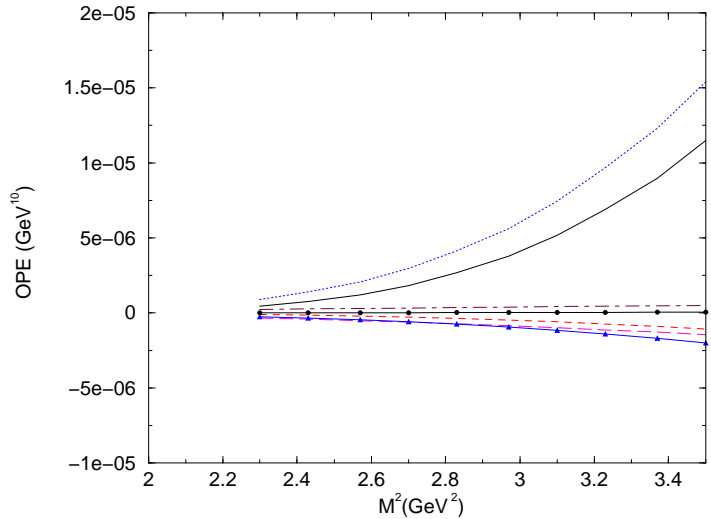,height=70mm}}
\caption{The OPE convergence for the sum rule for $Z_s$, using $\sqrt{s_0} =
5.0$ GeV. The solid with triangles, long-dashed, dashed, solide with circles,
dot-dashed and dotted lines give,
respectively, $m_s$ times the quark condensate, four-quark condensate,
gluon condensate,  $m_s$ times the mixed condensate, dimension eight condensate
and the perturbative contributions. The solid line gives the total OPE
contribution to the sum rule.}
\label{figzsope}
\end{figure}

The expressions obtained in Eqs.~(\ref{dim8}) for $\rho^{\qq^2}(s)$ and
$\Pi^{mix\qq}(q^2)$ should be changed to:
\beqa
&&\rho^{\qq^2}(s)=-{m_c^2\qq\ss\over 4\pi^2}\sqrt{1-4m_c^2/s},
\nn\\
&&\Pi^{mix\qq}(q^2)={m_c^2m_0^2\ss\qq\over 2^3\pi^2}\int_0^1
d\al{\al(1-\al)\over m_c^2-\al(1-\al)q^2}\left[1+{m_c^2
\over m_c^2-\al(1-\al)q^2}\right].
\label{scond}
\enqa

We get also two new contributions due to the strange quark mass:
\beqa
&&\rho^{m_s\qq}(s)={3m_s\over2^{4}\pi^4}
\int\limits_{\almi}^{\alma} {d\al\over\al}\left[{\ss\over4}
{(m_c^2-s\al(1-\al))^2\over1-\al}-m_c^2\qq\int\limits_{\bemi}^{1-\al}
{d\be\over\be}((\al+\be)m_c^2-\al\be s)\right],
\nn\\
&&\rho^{m_s\mix}(s)={m_sm_0^2\over 2^6\pi^6}\sqrt{1-4m_c^2/s}\left(
{\ss\over2}(2m_c^2-s)-3m_c^2\qq\right).
\label{ms}
\enqa

Using $m_s=(0.13\pm0.03)~\GeV$ \cite{sig}, and the continuum threshold in the
range
$\sqrt{s_0}=(5.1\pm0.1)~\GeV$ we see, from Fig.~\ref{figzsope}, that there is
a good OPE convergence for $M^2\geq2.5~\GeV^2$. From Fig.~\ref{figzsope} we
also see that, although proportional to $m_s$, the dimension four condensate
$m_s\qq$ (the solid line with triangles) is the most important condensate
contribution.

The upper limits for $M^2$ for each value of $\sqrt{s_0}$ are given in Table I.

\begin{center}
\small{{\bf Table I:} Upper limits in the Borel window for $Z_s$.}
\\
\begin{tabular}{|c|c|}  \hline
$\sqrt{s_0}~(\GeV)$ & $M^2_{max}(\GeV^2)$  \\
\hline
 5.0 & 2.80 \\
\hline
5.1 & 3.14 \\
\hline
5.2 & 3.43 \\
\hline
\end{tabular}\end{center}

\begin{figure}[h]
\centerline{\epsfig{figure=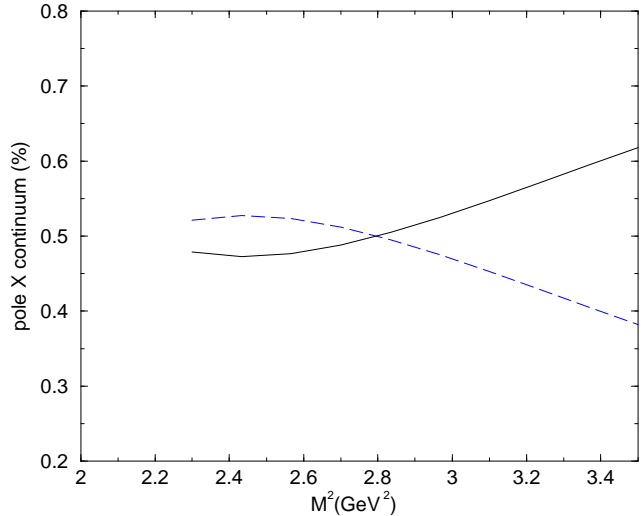,height=70mm}}
\caption{Same as Fig.~2 for  $Z_s$ using $\sqrt{s_0}=5.0~\GeV$.}
\label{figpvczs}
\end{figure}

In Fig.~\ref{figpvczs} we show the relative continumm (solid line) versus 
pole (dashed line) contribution, for $Z_s$ using $\sqrt{s_0}=5.0~\GeV$, from 
where we clearly see that the pole contribution is bigger than the continuum 
contribution for $M^2<2.80~\GeV^2$.

\begin{figure}[h]
\centerline{\epsfig{figure=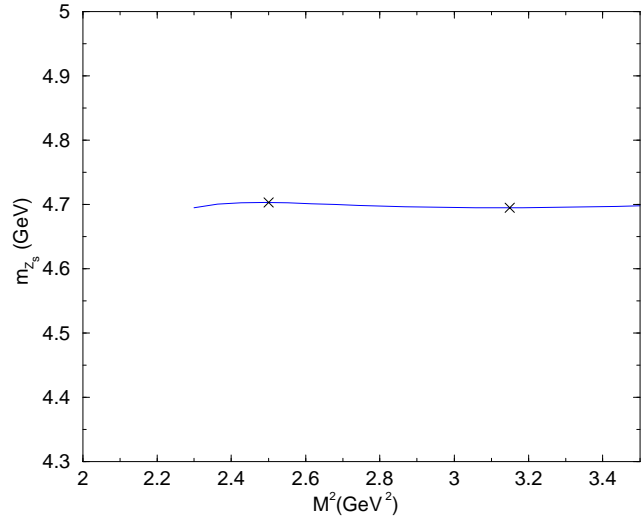,height=70mm}}
\caption{The $Z_s$ meson mass as a function of the sum rule parameter
($M^2$) for $\sqrt{s_0}=5.0$ GeV. The crosses delimit the region allowed for
the  sum rule.}
\label{figmzs}
\end{figure}

In the case of $Z_s$ we get a remarkable stability for the $Z_s$ mass,
in the allowed sum rule window, as a function of $M^2$
as can be seen by  Fig.~\ref{figmzs}.

Taking into account the variations on $M^2$, $s_0$, $m_s$ and $m_c$ in
the regions indicated above we get:
\beq
\lb{masszs}
m_{Z_{s}}=  (4.70\pm0.06)~\GeV~,
\enq
which is bigger than the $D_s^{*}D_1$ threshold $\sim4.5~\GeV$, indicating
that this state is probably a very broad one and, therefore, it might be very
dificult to be seen experimentally.


In conclucion,
we have presented a QCDSR analysis of the two-point
functions of the recently observed $Z^+(4430)$ meson. Due to  the closeness of
the $Z^+(4430)$ mass to the threshold of $D^{*+}(2010)\bar{D}_1(2420)$, we
have followed ref.~\cite{meng}, and have considered the $Z^+(4430)$ meson
as a $D^*D_1$ molecule. We have also presented a QCDSR study for the analogous
mesons $Z_{bb}$ and $Z_s$ considered as $B^*B_1$ molecule and $D^*_sD_1$ 
molecule respectively. We find very good OPE convergence for these three 
four-quark mesons, although this is not in general the case for tetraquark 
states \cite{tetra}. We got for $Z^+$ a mass in a very good agreement with the
experimental result.

In the case of $Z_s$ we have obtained a mass  bigger than the
$D_s^*D_1$ threshold. Therefore, our results indicate that
the  $Z_s$ meson is probably very broad.

\section*{Acknowledgements}
{This work has been partly supported by FAPESP and CNPq-Brazil,
and by the Korea Research Foundation KRF-2006-C00011.}


\end{document}